\documentclass[traditabstract]{aa} % for the abstract without structuration

\usepackage{graphicx}
\usepackage{amssymb}
\usepackage{txfonts}
\usepackage{mathrsfs}
\usepackage{array}
\usepackage{ulem}
\usepackage{natbib}
\bibpunct{(}{)}{;}{a}{}{,}   % to follow the A&A style
\begin{document}

\title{Milky  Way  demographics with  the  VVV survey\thanks{Based  on
    observations  taken  within  the  ESO  VISTA  Public  Survey  VVV,
    Programme ID 179.B-2002.}}

\titlerunning{Milky Way demographics with the VVV survey}

\subtitle{III. Evidence for a Great  Dark Lane in the 157 Million Star
  Bulge Color-Magnitude Diagram}

   \author{
  D.~Minniti\inst{1,2,3,4}
\and
  R.~K.~Saito\inst{5}
\and
  O.~A.~Gonzalez\inst{6}
\and
  M.~Zoccali\inst{2}
\and
  M.~Rejkuba\inst{7}
\and
  J.~Alonso-Garc\'{i}a \inst{2,3}
\and
  R.~Benjamin\inst{8}
\and
  M.~Catelan\inst{2,3}
\and
  I.~Dekany\inst{2,3}
\and
  J.~P.~Emerson\inst{9}
\and
  M.~Hempel\inst{2}
\and
  P.~W.~Lucas\inst{10}
\and
  M.~Schultheis\inst{11}
          }

\institute{  $^1$Departamento  de  Ciencias  F\'{i}sicas,  Universidad
  Andr\'es Bello, Rep\'ublica 220, Santiago, Chile\\
$^2$Instituto de Astronom\'{\i}a y Astrof\'{\i}sica, Pontificia
  Universidad  Cat\'{o}lica  de  Chile,  Av. Vicu\~na  Mackenna  4860,
  Casilla 306, Santiago 22, Chile \\
$^3$Millennium  Institute  of  Astrophysics, Av.  Vicu\~{n}a  Mackenna
  4860, 782-0436 Macul, Santiago, Chile\\
$^4$Vatican Observatory, Vatican City State V-00120, Italy\\
$^5$Universidade  Federal   de  Sergipe,  Departamento   de  F\'isica,
  Av.Marechal Rondon s/n, 49100-000, S\~ao Crist\'ov\~ao, SE, Brazil\\
$^6$European Southern Observatory, Casilla 19001, Santiago 19, Chile\\
$^7$European Southern Observatory, Karl-Schwarzschild-Strasse 2, D-85748 
Garching, Germany\\
$^8$Department of Physics, University  of Wisconsin, Whitewater, WI 53190,
USA\\
$^9$Astronomy Unit, School of Physics and Astronomy, Queen Mary University
of London, Mile End Road, London, E1 4NS, UK\\
$^{10}$Centre for Astrophysics Research, University of Hertfordshire, College
Lane, Hatfield AL10 9AB, UK\\
$^{11}$Universite de  Nice Sophia-Antipolis, CNRS,  Observatoire de Cote
d’ Azur, Laboratoire Cassiope e, 06304 Nice Cedex 4, France
}

   \date{Received ; accepted }

% \abstract{}{}{}{}{} 
% 5 {} token are mandatory
 
  \abstract{  The  new generation  of  IR  surveys  are revealing  and
    quantifying  Galactic   features,  providing  an   improved  $3-D$
    interpretation of  our own Galaxy.  We present an analysis  of the
    global distribution of dust clouds  in the bulge using the near-IR
    photometry  of  157  million   stars  from  the  VVV  Survey.   We
    investigate  the color-magnitude  diagram of  the Milky  Way bulge
    which shows  a red giant  clump of core  He burning stars  that is
    split in  two color  components, with a  mean color  difference of
    $(Z-K_{\rm   s})=0.55$    magnitudes   equivalent   to   $A_V=2.0$
    magnitudes.   We conclude that  there is  an optically  thick dust
    lane  at intermediate latitudes  above and  below the  plane, that
    runs   across  several  square   degrees  from   $l=-10^\circ$  to
    $l=+10^\circ$.    We   call   this   feature  the   ``Great   Dark
    Lane''. Although its exact distance  is uncertain, it is located in
    front of  the bulge.   The evidence for  a large-scale  great dark
    lane within the Galactic bulge  is important in order to constrain
    models of  the barred  Milky Way bulge  and to compare  our galaxy
    with external  barred galaxies, where these kinds  of features are
    prominent.   We discuss  two other  potential implications  of the
    presence of the Great Dark Lane for microlensing and bulge stellar
    populations studies.}

\keywords{Galaxy: center --- Galaxy: structure --- stars: late-type
  --- dust, extinction --- surveys}
\authorrunning{Minniti et al.}
\titlerunning{Milky Way Demographics with the VVV Survey III}

   \maketitle
%
%________________________________________________________________

\section{Introduction}
\label{sec:intro}

All-sky IR surveys, such as COBE and 2MASS, represent a revolution for
the  Galactic  structure  field.   These  surveys  have  provided  the
necessary tools to investigate  the large-scale structure of the Milky
Way galaxy. On  the other hand, the new generation  of IR surveys like
GLIMPSE             \citep{2005ApJ...630L.149B},            UKIDSS-GPS
\citep{2008MNRAS.391..136L},  and VVV  \citep{2010NewA...15..433M} are
now providing a much more detailed view of our Galaxy, identifying and
quantifying    structures    that    before    were    ambiguous    or
unclear. Particularly  important, is the capability  of recent surveys
to trace the red clump (RC)  stars across the Galaxy, even in the most
reddened regions.  RC stars are helium-burning  giants, counterpart of
the  metal-poor horizontal  branch  stars seen  in globular  clusters,
which have been proven  to be reliable distance indicators. Therefore,
by investigating the  distribution of RC stars we are  able to map the
morphology of the Galaxy \citep[e.g.][]{stanek+1994}.

The RC can  be easily identified in the  bulge color-magnitude diagram
(CMD)  and   its  mean  observed   magnitude  can  be   determined  by
constructing  the bulge  luminosity  function.  It  has recently  been
found     that     the    RC     in     the     bulge    is     double
\citep{2010ApJ...724.1491M,2010ApJ...721L..28N}. Both  RCs can only be
seen  simultaneously  along the  Bulge  minor  axis  fields and  their
magnitude difference becomes larger  as Galactic latitude increases. A
detailed mapping of  the distribution of RC stars  in the Bulge, using
2MASS   data,  allowed   \cite{2011AJ....142...76S}  to   confirm  the
interpretation  from  \cite{2010ApJ...724.1491M}  that the  double  RC
traces the X-shaped morphology of the bulge. Additionally, the stellar
kinematics     in     the    bright     and     faint    red     clump
\citep{2013A&A...555A..91V}, and the deeper  RC stars mapping based on
VVV   survey    data   \citep{wegg+2013}   further    confirmed   this
interpretation. 

Here we report  a different feature of the RC stars  in the bulge: the
observation  of a  split  on the  distribution  of RC  stars into  two
different   colors,   but  having   very   similar  $K_{\rm   s}$-band
magnitudes. The  small difference in magnitude is  consistent with the
color difference  being due to  extinction by dust. We  interpret that
this is  caused by a  dust lane that  runs across the whole  Milky Way
bulge  at intermediate  latitudes, a  feature  that we  will call  the
``Great Dark Lane'' of the Milky Way bulge.

This letter  is organized  as follows.  In  Section~\ref{sec:split} we
present the CMD for 157 million  stars in 300 sq.  deg.  of the bulge,
discussing the split  RC into two color components,  blue and red, and
the spatial distribution of the great dark lane. Section~\ref{sec:psf}
presents   a   further   detailed   exploration  of   one   field   at
$(l,b)=-0.6^\circ , -1.9^\circ$ that shows a mean color difference of:
$\Delta  A_{K_s}=1.79$ for  red  and blue  RC  components.  This  deep
pencil beam study  allows us also to explain why  this feature has not
been seen previously  in optical CMDs. Section~\ref{sec:test} presents
our interpretation of the small difference in magnitude of the RC seen
in our CMDs as  a geometric effect.  Finally, Section~\ref{sec:discus}
discusses some of the  implications for studies of galactic structure,
bulge stellar populations and microlensing experiments.

\section{The Color-Split in the Red Giant Clump of the Galactic Bulge}
\label{sec:split} 

Updating the  work of  \cite{2012A&A...544A.147S}, we have  merged all
$ZYJHK_{\rm    s}$    VVV   bulge    catalogs    appearing   in    the
CASU\footnote{http://casu.ast.cam.ac.uk/vistasp/}   database  as  v1.3
``Completed'' data, covering  300 sqdeg, from $-10^\circ <l<10^\circ$,
$-10^\circ <b<5^\circ$.  These come  from 196 bulge tiles that contain
239 million sources (with detection in any filter). Of these there are
157 million stellar sources, with stellar  flag in at least two of the
five   filters   (stellar   flag   denotes   good-quality   unblendend
sources). We  note that for the  10 tiles for  which the ``Completed''
data were not available, we  made use of ``Executed'', ``Aborted'' and
no flag data.

\begin{figure}[ht]
\includegraphics[bb=2.0cm 4.3cm 15cm 10cm,angle=-90,scale=0.57]{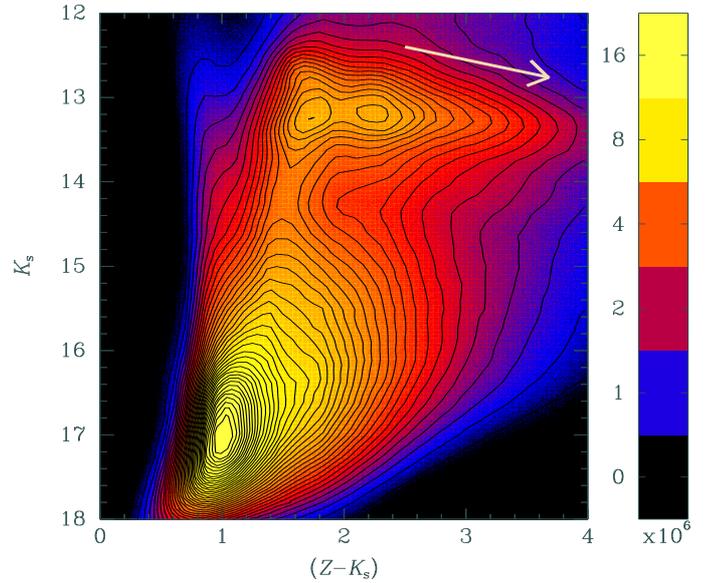}
 \caption{$K_{\rm s}$  vs. $(Z-K_{\rm s})$  color-magnitude diagram of
   66 million  stars from 300  sqdeg in the  bulge.  This Hess  CMD is
   very useful because it  reveals very faint features, the complexity
   of  the population,  the effects  of  reddening, and  the range  of
   magnitudes and colors spanning throughout the whole Milky Way bulge
   like never  before.  Notice in  particular the double shape  of the
   red giant  clump at $K_{\rm s}$~mag,  separated by $\Delta(Z-K_{\rm
     s})=0.55$~mag.  This  is due  to the great  dark lane  across the
   Milky Way bulge. Completeness in  the inner regions only becomes an
   issue for  $K_{\rm s}>16$ mag.  The reddening  vector is associated
   with an extinction of $E(B-V)=1$, based on the relative extinctions
   of the  VISTA filters, and  assuming the \cite{1989ApJ...345..245C}
   extinction law. Contour  lines mark density levels in  steps of 2\%
   from  the  maximum  density.   The  source  density,  in  units  of
   $10^6$\,sources mag$^{-2}$, is indicated in the vertical bar on the
   right.}
\label{fig:cmd}
\end{figure}

\begin{figure*}[ht]
\includegraphics[bb=8.5cm 0.5cm 17.2cm 16cm,angle=-90,scale=0.75]{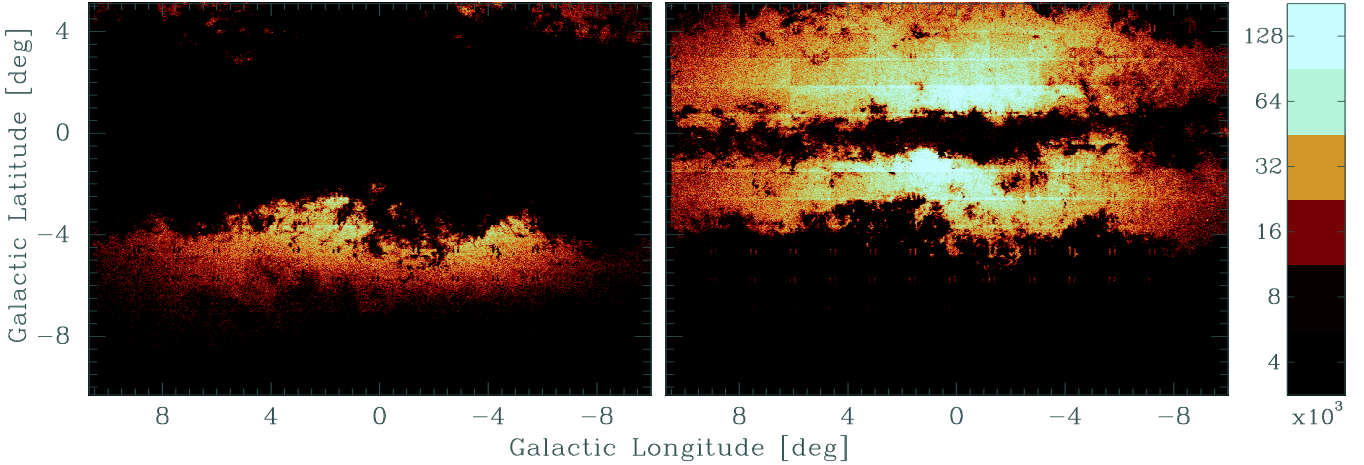}
 \caption{Left  panel: Spatial distribution  of stars  in the  blue RC
   with  $12.8<K_{\rm  s}<13.5$  and $1.5<(Z-K_{\rm  s})<2.0$.   Right
   panel:  Spatial   distribution  of  stars   in  the  red   RC  with
   $12.8<K_{\rm s}<13.5$  and $(Z-K_{\rm s})>2.1$.   These stars trace
   the location of the bulge dark lane. The innermost dark region with
   $-1<b<+1$ is  populated by  much more reddened  RC giants  and thus
   affected  by   completeness.  The  source  density,   in  units  of
   $10^3$\,sources deg$^{-2}$, is indicated in the vertical bar on the
   right.}
\label{fig:dust01}
\end{figure*}

The  resulting  $K_{\rm s}$  vs.   $(Z-K_{\rm  s})$  CMD is  shown  in
Fig.~\ref{fig:cmd}.   Only good-quality  (flag 1)  stellar  sources in
both  $Z$   and  $K_{\rm  s}$-band   filters  were  included   in  the
construction of  the CMD, which includes  a total of  about 70 million
sources.   These  unprecedented  VVV  data  allow us  to  analyze  the
large-scale  spatial  variations  of  this  CMD  and  the  effects  of
reddening \citep[e.g., ][]{2012A&A...544A.147S}.

The  effect of  the interstellar  extinction  on the  CMD is  evident,
spreading  the RC  stars along  the reddening  vector by  more  than 3
magnitudes   in   $(Z-K_s)$.   Although   this  effect   is   expected,
Fig.~\ref{fig:cmd}  shows that  this  re-distribution of  RC stars  in
color is not continuous but, instead, presents two clear overdensities
that split the RC into blue and red components (we use this convention
to avoid confusion).

The mean  observed $(Z-K_{\rm s})$ color difference  between these two
RCs is  $0.55\pm0.03$~magnitudes.  If this color  difference is purely
due to  reddening, this is equivalent to  $E(B-V)=0.65$ and $A_V=2.01$
~magnitudes, using the ratios $E(Z-K_{\rm s}) = 1.18\times E(B-V)$ and
$E(J-K_{\rm    s})=0.50\times    E(B-V)$   \citep{2011COAS…2011.145P},
corresponding to a  very dark feature indeed.  A  similar double color
RC  is  observed   in  the  $K_{\rm  s}$  vs.    $(J-K_{\rm  s})$  CMD
\citep[e.g., ][]{2012A&A...544A.147S}, but  the separation between the
blue  and   red  RC  is   best  appreciated  with  the   longer  color
\textit{baseline} provided by the $(Z-K_{\rm s})$ color.

In  order  to   evaluate  this  and  to  map   the  different  spatial
distribution of  these sources, we adopt the  following boundaries for
the two components of the red  clump (RC): a blue RC with $12.8<K_{\rm
  s}<13.5$ and $1.5<(Z-K_{\rm s})<2.0$, and a red RC with $12.8<K_{\rm
  s}<13.5$ and $(Z-K_{\rm s})>2.1$.   The spatial distribution of blue
and   red   components   of   the   RC   are   shown   separately   in
Fig.~\ref{fig:dust01}. We notice that  the spatial distribution of the
red  component has  nearly specular  edges with  respect to  the bluer
RC. Figure~\ref{fig:dust01} shows that  the great dark lane is present
across the  whole bulge, above and  below the plane.  We  see that the
distribution of  stars affected by the  dark lane is  coherent and not
patchy,  with  a  sharp  transition  at  latitudes  $|b|<4^\circ$  and
extending  for  many  square   degrees  in  Galactic  longitude,  from
$l=-10^\circ$  to $l=+10^\circ$. Unfortunately,  although the  VVV maps
extend  in longitude  from $l=350^\circ$  to $l=295^\circ$,  they only
cover  a  narrow strip  along  the  plane  for latitudes  $-2.25^\circ
<b<+2.25^\circ$ thus  not allowing us to investigate  its extension in
longitude.

Assuming  a   smooth  stellar  density  distribution   for  the  bulge
\citep{wegg+2013}, the  marked split in the color  distribution of the
RC suggests that  there must be a region in the  bulge where stars are
affected by  a dust feature that  is optically much  thicker than that
affecting  the stars  in the  blue RC  component.  We  would otherwise
expect a smooth color transition between RC stars for regions that are
differently affected by extinction.  Furthermore, if this dust feature
was located  in the middle  of the bulge  population, we would  see RC
stars from the blue and red  components mixed across the bulge. On the
contrary,  we see that  both components  are spatially  distributed in
specific regions of  the bulge.  These properties lead  us to conclude
that there  is an  optically thick dust  feature, with  sharp, clearly
marked edges that  is located \textit{in front} of  the bulge. We name
this dust feature  the ``Great Dark Lane" of the  Milky Way bulge.  We
note that  according to  the Galactic latitudes  up to which  the dark
lane extends  ($|b|<4^\circ$), the projected distance  from the plane,
at a  mean distance of 6~kpc (i.e., at  the near side of  the bulge),
would be of $\sim$400~pc.

\section{PSF Photometry of a Field at ($l,b = -0.6^\circ, -1.9^\circ$)}
\label{sec:psf}

\begin{figure*}[ht]
\includegraphics[bb=6.6cm -2cm 15.5cm 9cm,angle=-90,scale=0.58]{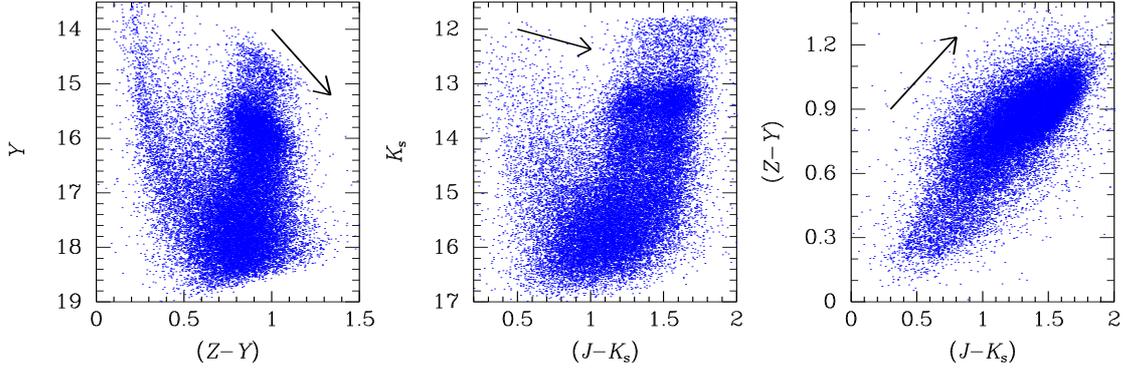}
 \caption{Data from PSF photometry. Left:  $Y$ vs.  $(Z-Y)$ CMD of one
   chip  of the  tile b305  centered at  $(l,b)=-0.6,-1.9$  showing no
   double  clump  at  more  optical  wavelengths  due  to  extinction.
   Middle:  $K_{\rm s}$  vs $(J-K_{\rm  s})$  CMD for  the same  stars
   showing double clump due to  extinction.  This is the observed CMD,
   it  is  not  dereddened.    Right:  $(Z-Y)$  vs.   $(J-K_{\rm  s})$
   color-color   diagram  for  stars   with  clump   giant  magnitudes
   $13.0<K_{\rm  s}<13.6$.  The  reddening  vector in  each  panel  is
   associated with an extinction  of $E(B-V)=2$, based on the relative
   extinctions    of   the   VISTA    filters,   and    assuming   the
   \cite{1989ApJ...345..245C} extinction law.}
\label{fig:psf}
\end{figure*}

The CASU photometry is aperture  photometry, it then may be limited by
crowding in the  inner regions, even though the  effect is expected to
be minor  for stars  at the  RC level and  more important  for fainter
stars. In order to confirm the reality of this great dark lane feature
and  quantify the  effect produced,  we decided  to obtain  deeper PSF
photometry in a representative inner field -- VVV tile b305 -- located
at  $(l,b)=-0.6^\circ,-1.9^\circ$.   The  right  and  middle  panels  of
Fig.~\ref{fig:psf} show the deep  $K_{\rm s}$ vs.  $(J-K_{\rm s})$ and
$Y$ vs.   $(Z-Y)$ CMDs  for one  chip of this  tile obtained  with PSF
photometry  using  DoPhot \citep{1993PASP..105.1342S,2012AJ..143.70A}.
This field was selected because of its straightforward interpretation:
it crosses  the edge of the  dust lane containing stars  for both blue
and red RC components.

Again, the $K_{\rm s}$ vs. $(J-K_{\rm  s})$ RGB in this field is split
in  two  well-defined branches,  and  two  red  giant clumps  can  be
identified. The  separation can  be easily measured  in the  red giant
clump  split in  color by:  $\Delta(J-K_{\rm  s})=0.30\pm0.03$~mag. If
this color jump is interpreted as only due to the effect of extinction
from an intervening cloud along  the line of sight, this is equivalent
to  $\Delta(B-V)=0.57\pm0.06$, and $\Delta(A_V)=1.78$.   These figures
are  similar  to  the  ones  obtained  using  the  whole  $K_{\rm  s}$
vs. $(Z-K_{\rm s})$ CMD in Section~\ref{sec:split}.

In contrast, the  disk main sequence in this field  is well defined in
the  $K_{\rm  s}$ vs.  $(J-K_{\rm  s})$  CMD,  somewhat broad  due  to
foreground differential reddening, but definitely not bimodal like the
RGB, indicating that these disk stars lie
predominantly in the foreground of the great dust lane..

The mean  reddening for  the region  located at $-0.68  < l  < -0.52$,
$-1.98 < b  < -1.82$ was measured by  \cite{2011A&A...534A...3G} to be
$E(B-V)=1.46$  using  the  \cite{1989ApJ...345..245C}  reddening  law.
There is evidence  that the reddening law varies  in the inner regions
\citep[e.g.,][]{2009ApJ...696.1407N},   but   choosing   a   different
reddening law does not affect  the present results.  Taking this value
as mean reddening, we find  that the extinction varies from an average
of $E(B-V)=1.18$  for the bulge stars  not affected by  the great dark
lane to an  average of $E(B-V)=1.75$ for the  stars located behind it.
This total extinction applies only to this particular field b305. Even
though we  see the coherent split  of the RGB across  the whole bulge,
this may vary  in distance and a finer 3-D  mapping for the extinction
is warranted  (e.g., with  individual RR Lyrae  discovered by  the VVV
survey).

Many  of  the  inner  fields  have been  mapped  by  the  microlensing
experiments              in              optical             passbands
\citep{2000ApJ...541..734A,2002AcA....52..217U}.    However,   such  a
split clump has never been reported, which seems in contradiction with
the present findings. The reason for that becomes evident in the bluer
$Y$ vs. $(Z-Y)$  CMD, that is closer to optical  (MACHO, OGLE) CMDs of
the same stars of Fig.~\ref{fig:psf},  left panel.  The RGB in the $Y$
vs. $(Z-Y)$  CMD is  very broad  and the red  giant clump  is extended
along the direction of the  reddening vector, but at these wavelengths
the differential  extinction blurs the  split color of the  RGB, which
instead appears as a single wide  branch, with a color spread of about
$0.5$~mag.  This color scatter is real and not an effect due to larger
photometric errors  in the $ZY$  passbands, because the  photometry in
all these bands  is better than $0.02$~magnitudes at  the level of the
red giant clump.

In the  right panel of  Fig.~\ref{fig:psf} we selected the  stars with
$K_{\rm s}$ magnitudes that include  the RC as $ 13.0<K_{\rm s}<13.6$,
and plot  the color-color  diagram for this  bulge field in  the right
panel  of Fig.~\ref{fig:psf}.   The  stars are  aligned following  the
reddening vector. The red clump  bimodality is in $(J-K_{\rm s})$, but
not in $(Z-Y)$,  and at the same time the  foreground main sequence is
unimodal.   This explains  why this  was not  noticed in  the previous
optical CMDs that  mapped the inner bulge region,  such as for example
the  photometry of  the microlensing  experiments.  Fig.~\ref{fig:psf}
shows the  advantage of  having $ZYJHK_{\rm s}$  passbands in  the VVV
survey:  there are  features  in the  CMDs  that can  be ambiguous  or
missing when we inspect only single CMDs.

\section{The geometric effect and metallicity}
\label{sec:test}

An interesting aspect seen in  Fig.~\ref{fig:cmd} is that the blue and
red RCs  have similar $K_{\rm s}-$band magnitude,  with a distribution
flatter than that expected by the corresponding reddening vector.

The mapping of  the RC stars across the Milky Way  bulge has been used
by several authors  in order to trace the geometry of  the bar and its
splitting      in      a      X-shaped     structure      \citep[e.g.,
][]{2012ApJ...744L...8G,      2011AJ....142...76S,2012A&A...544A.147S}.
These  results  demonstrate that  the  RC  magnitude  varies with  the
longitude across  the bulge in  the sense that at  positive longitudes
the RC appears brighter than at negative longitudes.  Moreover, the RC
magnitude also varies with  the Galactic latitude, since the projected
distance of the Galactic bar changes when measured at higher values of
$b$ \citep[e.g.,][]{2011AJ....142...76S,2012ApJ...744L...8G}.

At a given value of Galactic  longitude, the red RC, which is affected
by the dust lane, is constrained to regions at low $|b|$. On the other
hand, the  blue RC  which is  unaffected by the  dust lane  comes from
regions at higher $|b|$ (further  from the plane). Thus, the magnitude
of the clumps vary due to the projected distance of the bar -- at high
$|b|$ (responsible to  the blue RC) the bar  is projected farther than
in the Galactic plane (the red  RC, at low $|b|$). This effect is more
significant at negative longitudes and prevents the blue and red RC to
follow the reddening vector.

\begin{figure}[ht]
\includegraphics[bb=0cm 7.5cm 10cm 15cm,angle=-90,scale=0.75]{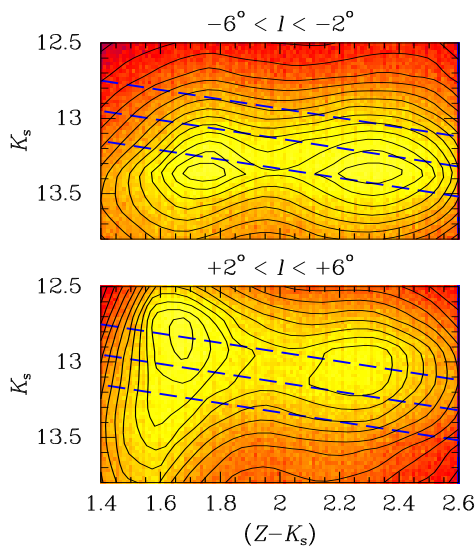}
 \caption{$K_{\rm  s}$ vs.   $(Z-K_{\rm s})$  color-magnitude diagrams
   for the region around the red  clump position.  In the top panel we
   show data  for $-6^\circ<l<-2^\circ$ and  in the bottom  panel data
   for the  region $+2^\circ<l<+6^\circ$.   The dashed lines  mark the
   slope of reddening vectors based on the relative extinctions of the
   VISTA   filters,   and   assuming  the   \cite{1989ApJ...345..245C}
   extinction law.  Contour lines mark  density levels in steps of 5\%
   from the maximum density.}
\label{fig:test}
\end{figure}

We clarify this  effect by slicing two bulge  regions, one at positive
longitudes  $+2^\circ<l<+6^\circ$  and  another  symmetric  region  at
negative longitudes ($-6^\circ<l<-2^\circ$). Fig.~\ref{fig:test} shows
the $K_{\rm s}$ vs.  $(Z-K_{\rm s})$ CMDs for the region around the RC
position for both regions. While  the difference in colour between the
blue and red RC for both  regions is consistent with the presence of a
thick dust lane, the position in magnitude changes in all cases due to
the projection effect.

If the bimodal RGB and RC are not due to extinction, we have to search
for alternative  explanations. One possibility is the  presence of two
stellar populations with  different metallicities.  The large majority
of the stars in the bulge have metallicities of $-0.7<$~[Fe/H]~$<+0.2$
\citep[e.g.,][]{2008A&A...486..177Z,2013A&A...552A.110G}. Based on the
isochrones of \cite{2000A&AS..141..371G},  the mean absolute magnitude
of red clumps stars  varies from $M_K=-1.306$~mag ([Fe/H]~$=-0.68$ and
$Z=0.004$) to $M_K=-1.571$~mag ([Fe/H]~$=+0.2$ and $Z=0.03$).

In  principle since all  these sters  are present  in all  fields, the
difference  in   metallicity  produces  a  {\it   spread}  in  $K_{\rm
  s}$-magnitude  of the  RC (width)  of  the order  of $\Delta  K_{\rm
  s}\sim0.27$~mag.  However, in order to  produce a {\it shift} in the
mean magnitude of $\sim0.27$~mag in  $\Delta K_{\rm s}$ one would need
a shift  in the mean magnitude  of the RC  stars in the redder  and in
bluer clump of the order of $\sim1$~dex.

\cite{2013A&A...552A.110G} have  recently produced a  full metallicity
map for the Galactic bulge.  The $l<0^\circ$ region is more metal-poor
and  for that  region  the average  magnitude  of the  clump would  be
fainter than  for $l>0^\circ$. However, the  red RC, which  is the one
affected  by the  dust lane,  is constrained  to  $|b|<3^\circ$.  Both
$l>0^\circ$  and   $l<0^\circ$  parts   of  the  bulge   have  similar
metallicity for $|b|<3^\circ$.  The RC  magnitude of the red RC should
therefore  depend purely  on  the distance  and  extinction.  Given  a
similar  extinction  in  the  positive  and  negative  $l$,  the  only
difference in  $\Delta K_{\rm s}$ of the  red RC for $+l$  and $-l$ is
then due to distance/projection.

In  absence of  extinction,  purely due  to  sampling different  $|b|$
(assuming that the only difference is metallicity), the blue RC should
be fainter than the red one. The difference in metallicity between the
positive and negative latitude  of the bulge between $|b|<3^\circ$ and
$|b|>3^\circ$ for $l>0^\circ$ is $\Delta$[Fe/H]$\sim0.2$ and is larger
$\Delta$[Fe/H]$\sim0.4-0.5$  for  $l<0$. It  could  contribute to  the
different slope between the the blue and red RC. 

The value of the shift in $\Delta K_{\rm s}$ on the blue RC (comparing
to red  RC) due to a  metallicity effect only  would be $\sim0.08$~mag
for $l>0^\circ$ and $\sim0.13$~mag for $l<0^\circ$.  Therefore, in the
full  map there  is a  mixture from  both the  projection/distance and
metallicity,  but  the metallicity  effect  is  smaller.  The  maximum
difference in the mean metallicity  produces a shift in $\Delta K_{\rm
  s}\sim0.1$~mag.

Similar  analysis can  be applied  to  the $Z$-band,  and the  results
demonstrate that  the maximum $\Delta(Z-K_{\rm  s})$ colour difference
due to metallicity is smaller than $\sim0.2$~mag.  Thus the difference
in  metallicity is  not enough  to account  for the  split of  the two
RCs. Additional double  populations made of two different  ages can be
discarded using similar arguments, and  also because we do not see the
corresponding main  sequence of a younger  population.  The extinction
caused by an optically thick dust  lane seems to be the only remaining
explanation for the observed split RC in color.

\section{Discussion and Implications}
\label{sec:discus}

The  near-IR CMDs  of  157 million  sources  across 300  sqdeg in  the
Galactic  bulge  shows  that  the   red  clump  giants  are  split  in
color.  Selecting the sources  belonging to  the blue  and red  RCs we
study the global distribution of  dust in the Milky Way bulge, finding
evidence for a large coherent dust absorption across the bulge that we
call the great dark lane.

Even though the distance is unknown, we suggest that this dust lane is
probably located  in front of the  bulge. his is supported  by the 3-D
dust extinction  maps by \cite{2014A&A...566A.120S}, who  find a large
jump in  extinction at low latitude  located at a distance  of about 6
kpc along the line of sight to the bulge. This may be the signature of
a  dust ring  as  seen in  other  galaxies that  have suffered  recent
mergers (such as M64 or NGC5128).

The  discovery  of this  “great  dark lane”  at  low  latitudes is  of
particular interest  for studies  of Galactic structure.  The existing
literature does  not contain  reports of such  a large  extension dark
lane (in scale height and across the whole bulge), and detailed maps
and  modelling are  needed in  order to  test this  important galactic
feature.

\cite{1992MNRAS.259..345A}  studied  and  modelled the  existence  and
shapes of gas/dust in barred bulges, and showed the presence of shocks
with velocity jumps and  characteristic shapes. These large-scale dust
lanes are located at both sides  of the bar and leading (as opposed to
the spiral arms  that are trailing). The characteristics  of the bulge
great dark lane observed here should be tested against the predictions
of    models    of    barred    galaxies   like    the    Milky    Way
\citep{1992MNRAS.259..345A,1999A&A...345..787F,2006A&A...455..963R,
2008A&A...489..115R,2008MNRAS.389..545M,2010PASJ...62.1413B}.

This  result is  interesting  for the  interpretation of  microlensing
events, where  microlensing sources in  front of the bulge  great dark
lane  are brighter  in optical  passbands  than those  behind. In  one
specific field  we found  that the difference  in magnitudes  would be
$\Delta_V = 1.8\,mag$, and depending on the distance to the great dark
lane,  such  large difference  would  bias  against  the detection  of
microlensing source stars  that are located behind it,  which would in
turn have some impact in microlensing optical depth calculations.  The
fields   where   most   microlensing   events   have   been   detected
\citep[e.g.,][]{2000AcA....50....1U,2004AAS...205.2803T,2005ApJ...631..879P}
lie in the direction where we see the bulge great dark lane.

Additional interest  for understanding the  effect of the  bulge great
dark lane arises when trying to  estimate the age of the bulge stellar
population. Traditionally ages have been determined using deep CMDs of
clear bulge  windows (e.g. Baade´s  window, SWWEPS field),  that yield
old                                                                ages
\citep[e.g.,][]{2003A&A...402..565O,2011ApJ...735...37C,2013arXiv1309.4570V},
but the measurement of ages  for inner fields have been complicated by
increased extinction and crowding.

Clearly, the bulge Great Dark  Lane presented here should be mapped in
greater  detail in  order to  understand its  distance,  geometry, and
effects on stellar population  and Galactic structure studies, as well
as microlensing studies.

\begin{acknowledgements}

We  gratefully acknowledge  use of  data  from the  ESO Public  Survey
programme  ID 179.B-2002  taken  with the  VISTA  Telescope, and  data
products from the Cambridge Astronomical Survey Unit, and funding from
the  BASAL-CATA Center  for Astrophysics  and  Associated Technologies
PFB-06, the Milky Way Millennium  Nucleus from the Ministry of Economy
ICM grant  P07-021-F, Proyecto FONDECYT Regular 1130196,  and from the
National   Science  Foundation  under   Grant  PHYS-1066293   and  the
hospitality of the  Aspen Center for Physics. We  warmly thank the ESO
Paranal  Observatory staff  for performing  the observations  and Mike
Irwin, Eduardo  Gonzalez-Solares, and Jim  Lewis at CASU  for pipeline
data processing support. R.K.S.  acknowledges support from CNPq/Brazil
through projects  310636/2013-2 and 481468/2013-7.  M.Z.  acknowledges
support  by  Proyecto FONDECYT  Regular  1110393.  J.A.   acknowledges
support by Proyecto FONDECYT Postdoctoral 3130552.

\end{acknowledgements}

\renewcommand*{\bibfont}{\tiny} \bibliographystyle{aa}

\bibliography{mybiblio}

\end{document}